\documentclass[showpacs,aps,prd,twocolumn,nofootinbib]{revtex4}
\usepackage{amsmath}
\usepackage{epsfig}
\usepackage{subfigure}
\usepackage{float}
\usepackage{longtable}

\newcommand{\ep}{\varepsilon}

\newcommand{\eqs}[1]{\begin{equation} \begin{split} #1\end{split} \end{equation} }

\newcommand{\ks}[1]{#1 \!\!\! \slash } 

\newcommand{\ga}{\gamma^5}

\newcommand{\ie}{{\it i.e.}}
\newcommand{\eg}{{\it e.g.}}
\newcommand{\etal}{{\it et al.}}

\newcommand{\ce}[1]{Eq.~(\ref{#1})}
\newcommand{\ced}[2]{Eqs.~(\ref{#1}) \& (\ref{#2})}
\newcommand{\cf}[1]{{Fig.~\ref{#1}}}
\newcommand{\ct}[1]{{Table~(\ref{#1})}}

\newcommand{\nn}{\nonumber}

\newcommand{\beq}[1]{
\begin{equation}\label{#1}}
\newcommand{\eeq}{\end{equation}}
\newcommand{\bea}[1]{
\begin{eqnarray}\label{#1}}
\newcommand{\eea}{\end{eqnarray}}

\newcommand{\eq}[1]{\begin{equation}#1\end{equation} }
\newcommand{\out}{\raise-3pt\hbox{\scriptsize    out}}

\begin{document}

\title{Hard exclusive electroproduction of a pion in the backward region}

\author
{J.P. Lansberg$^{a}$, B. Pire$^{a}$ and  L. Szymanowski$^{a,b,c}$
}

\preprint{CPHT--  hep-ph/yymmnnn}

\affiliation{
$^{a}$Centre de Physique Th\'eorique, \'Ecole Polytechnique, CNRS, 
91128 Palaiseau, France
\\$^{b}$Physique Th\'eorique Fondamentale, Universit\'e de Li\`ege,
17 All\'ee du 6 Ao\^ut, B\^atiment B5a, B-4000 Li\`ege-1, Belgium\\
$^{c}$Soltan Institute  for   Nuclear  Studies,  Warsaw,   Poland 
}

\begin{abstract}
We   study   the   scaling  regime   of  pion electroproduction
in the backward region,  $e N  \to e' N' \pi $. We compute the leading-twist amplitude
in the kinematical region, where it factorises into a short-distance matrix element and 
 long-distance dominated nucleon Distribution Amplitudes  and nucleon to pion Transition Distribution 
Amplitudes. Using the chiral limit of the latter, we obtain a first 
estimate of the cross section, which may be experimentally studied  at JLab or Hermes.
\end{abstract}
\maketitle

\section{Introduction}

In~\cite{TDApiproton,Pire:2005mt}, we introduced the framework to study
backward pion electroproduction
\eq{\gamma^\star(q) N(p_1)  \to N'(p_2) \pi(p_\pi),}
on a proton (or neutron) target, 
in the Bjorken regime ($q^2$ large and $q^2/(2 p_1.q)$ fixed) in terms of Transition
Distribution Amplitudes (TDAs), 
 as well as the reaction
$N(p_1) \bar N (p_2) \to \gamma^\star(q) \pi(p_\pi)$
in the  near forward region. This extended the concept of 
Generalised Parton Distributions. 
Such an extension of the GPD framework has already
been advocated in the pioneering work of~\cite{Frankfurt:1999fp}.

The TDAs involved in the description of Deeply-Virtual Compton Scattering (DVCS) 
in  the backward kinematics 
\eq{ \gamma^\star (q) N(p_1)  \to N'(p_2) \gamma(p_\gamma) }
and the reaction
$N(p_1) \bar N (p_2) \to \gamma^\star(q) \gamma(p_\gamma)$ 
in the near forward region were given  in~\cite{Lansberg:2006uh}.

This followed the same lines as in~\cite{TDApigamma}, where we have argued that 
factorisation~theorems~\cite{Collins:1996fb}
for exclusive processes apply to the case of the reaction
$\pi^-\,\pi^+ \, \to \, \gamma^*\,\gamma$
in the kinematical regime where the off-shell photon is highly virtual (of 
the order of the energy squared of the reaction) but the momentum transfer $t$
is small.  Besides, in this simpler mesonic case, a perturbative limit may be  
obtained~\cite{Pire:2006ik}
for the $\gamma^\star$  to $\rho$ transition. For the 
$ \gamma \to \pi$ one, 
we have recently shown~\cite{TDApigamma-appl} that experimental analysis of processes such
as $\gamma^\star \gamma \to \rho \pi$ and  $\gamma^\star \gamma
 \to \pi \pi$,  which involve the latter 
TDAs, could be carried out,~\eg~the background from the Bremsstrahlung
is small if not absent and rates are sizable at present $e^+ e^-$ facilities.

Whereas in the pion to photon case, models used for 
GPDs~\cite{Tiburzi:2005nj,Broniowski:2007fs,noguera,GPD_pion} 
could be applied to TDAs since they are defined from matrix elements of the same quark-antiquark operators, 
this is not obvious for the nucleon to meson or photon TDAs, which are defined from matrix elements of 
a three-quark operator. Before estimates based on models such as the meson-cloud 
model~\cite{Pasquini:2006dv} become available, it is important to use model-independent information 
coming from general theorems. We will use here the constraints for the proton to pion TDAs derived 
in the chiral limit.

The structure of this paper is the following: first, we recall the necessary kinematics
related to hard electroproduction of a pion as well as the definitions of the proton to pion 
TDAs, which enter the description of the latter
process in the backward region; secondly, we establish the limiting constraints on the 
TDAs in the chiral limit when the final-state pion is soft; thirdly, we calculate the hard
contribution for the process; hence, extrapolating the limiting value of the TDAs to the large-$\xi$ 
region, we give a first evaluation of the unpolarised cross section, 
by restricting the analysis of the hard part 
to the sole Efremov-Radyushkin-Brodsky-Lepage (ERBL) region, where all the three quarks struck 
by the virtual photon have
positive light-cone momentum fraction of the target proton.

This analysis is motivated
by the experimental conditions~\cite{Laveissiere:2003jf,CLAS,Airapetian:2001iy}
of JLab and Hermes at moderate electron energies.
Related processes with three-quark exchanges in the hard scattering
were recently studied in~\cite{Braun:2006td} similarly to what was proposed
in~\cite{Pobylitsa:2001cz}.

\section{Kinematics and definitions}

\subsection{The electroproduction process $e P \to e' P'\pi^0$}

Let us first
recall the kinematics for the electron proton collisions (see \eg~\cite{Mulders:1990xw}).
As usual, we shall
work in  the one-photon-exchange approximation and consider the differential cross section
for $  \gamma^\star(q) P(p_1)  \to P'(p_2) \pi^0(p_\pi) $ 
in the center-of-mass frame of the pion and the final-state proton (see the kinematics in
\cf{fig:kinematics-bEPM}). The photon flux 
$\Gamma$ is defined in the Hand convention to be 
\begin{equation}
\Gamma=\frac{\alpha_{em}}{2\pi^2}\ \frac{E_{e'}}{E_e}\ \frac{W^2-M^2}{2 M Q^2}
\ \frac{1}{1-\epsilon},
\end{equation}
with $E_e$ the energy of the initial electron in the lab frame (beam energy), $E_{e'}$
the one of the scattered electron, $W$ the invariant mass of the
$P'\pi^0$ pair, $M$ the proton mass, $Q^2$ the virtuality of the exchanged photon
($Q^2=-q^2=-(p_e-p_{e'})^2$) and $\epsilon = \frac{1-y}{1-y+y^2/2}$  ($y = \frac{q.p_1}{p_e.p_1}$)
 its linear polarisation parameter. 
The five-fold differential 
cross section for the process $e P \to e' P'\pi^0$ can then be reduced
to a two-fold one, expressible in the center-of-mass frame of the $P'\pi^0$ pair, times the flux
factor $\Gamma$:
\begin{equation}
\frac{d^5\sigma}{dE_{e'}d^2\Omega_ed^2\Omega_{\pi}^\ast}=
\Gamma\ \frac{d^2\sigma}{d^2\Omega_{\pi}^\ast},
\end{equation}
where $\Omega_e$ is the differential solid angle for the scattered electron
in the lab frame, $\Omega_{\pi}^*$ is the differential solid angle for the
pion in the $P'\pi^0$ center-of-mass frame, such that $d\Omega_{\pi}^*=d\varphi \, d \cos \theta^*_\pi$.
$\theta^*_\pi$ is defined as the polar angle between the virtual photon and the pion in the latter system.
$\varphi$ is the azimuthal angle between the electron plane and the plane of the 
process $\gamma^\star P \to P' \pi^0$ (hadronic plane)
($\varphi=0$ when the pion is emitted in the half plane containing the outgoing electron, see
also~\cf{fig:kinematics-bEPM}).

\begin{figure}[h]
\includegraphics[height=5cm]{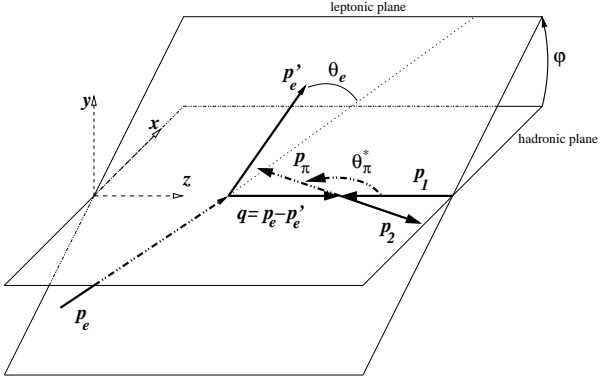}
\caption{(Backward) electroproduction of a pion on a proton in the center-of-mass frame of the
$\gamma^\star$ proton.}
\label{fig:kinematics-bEPM}
\end{figure}

In general, we have contributions from different polarisations of the photon. For that
reason, we define four polarised cross sections, which do not depend on $\varphi$ but only
on $W$, $Q^2$ and $\theta^*_\pi$, $d^2\sigma_{\mbox{\tiny T}}$,
$d^2\sigma_{\mbox{\tiny L}}$, $d^2\sigma_{\mbox{\tiny TL}}$ and
$d^2\sigma_{\mbox{\tiny TT}}$. The $\varphi$ dependence is therefore more explicit since
\begin{eqnarray}
\frac{d^2\sigma}{d\Omega_{\pi}^*} &=&
\frac{d^2\sigma_{\mbox{\tiny T}}}{d\Omega_{\pi}^*}\ +
\ \epsilon\ \frac{d^2\sigma_{\mbox{\tiny L}}}{d\Omega_{\pi}^*}
+ \sqrt{2\epsilon(1+\epsilon)}\ \frac{d^2\sigma_{\mbox{\tiny TL}}}
{d\Omega_{\pi}^*}\cos{\varphi} \nonumber \\
& & +\ \epsilon\ \frac{d^2\sigma_{\mbox{\tiny TT}}}{d\Omega_{\pi}^*}\cos{2\varphi}.
\label{sigma_development}
\end{eqnarray}

As we shall show below, at the leading-twist accuracy, the QCD mechanism considered here contributes only 
to $\frac{d^2\sigma_{\mbox{\tiny T}}}{d\Omega_{\pi}^*}$ and 
$\frac{d^2\sigma_{\mbox{\tiny TT}}}{d\Omega_{\pi}^*}$.

\subsection{The subprocess $\gamma^\star P \to P'\pi^0$}

In the scaling regime, the amplitude for
$  \gamma^\star P(p_1)  \to P'(p_2) \pi(p_\pi) $
 in the backward kinematics --namely small  $u=(p_\pi -p_1)^2=\Delta^2$ or $\cos \theta^*_\pi$ 
close to -1-- then involves the  TDAs $T(x_{i}, \xi, \Delta^2)$, 
where $x_i$ ($i=1,2,3$) denote the light-cone-momentum fractions carried by participant 
quarks and $\xi$ is the skewedness parameter such that $2\xi =x_1+x_2+x_3$.

The amplitude is then a convolution of the proton DAs, a 
perturbatively-calculable-hard-scattering amplitude and the TDAs,
defined from the Fourier transform of a matrix element of a three-quark-light-cone operator between a 
proton and a meson state.
We have shown  that these TDAs obey QCD evolution equations,
which follow from the renormalisation-group equation of  the
three-quark operator. 
Their $Q^2$ dependence is thus completely  under control.

\begin{figure}[h]
\includegraphics[height=6.5cm]{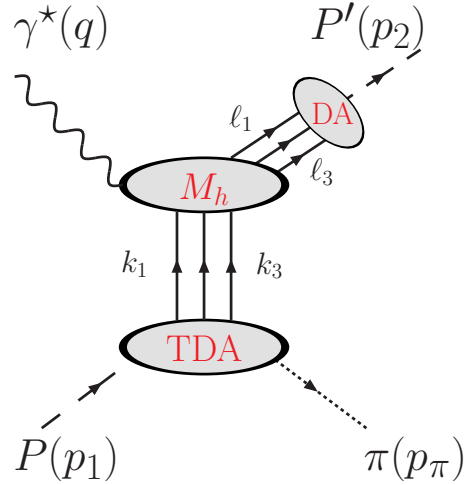}
\caption{The factorisation of the process $\gamma^\star P\to P'\;\pi$ into
proton-distribution amplitudes (DA), the hard-subprocess amplitude ($M_h$) and 
proton $\to$ pion transition distribution amplitudes (TDA) .
 }
\label{fig:fact-ampl}
\end{figure}

The momenta of the process $\gamma^\star P \to P' \pi $ are defined as shown in \cf{fig:kinematics-bEPM}
and \cf{fig:fact-ampl}.
The $z$-axis is chosen along the initial nucleon and the virtual photon  momenta
and the $x-z$ plane is identified 
with the collision or hadronic plane (\cf{fig:kinematics-bEPM}). Then, we define the 
light-cone vectors $p$ and $n$ ($p^2$=$n^2$=0) 
such that $2~p.n=1$, as well as $P=\frac{1}{2} (p_1+p_\pi)$, $\Delta=p_\pi -p_1$ and its 
transverse component $\Delta_T$ ($\Delta_T.\Delta_T=\Delta_T^2<0$), which we choose to be along 
the $x$-axis. From those, we define  $\xi$ in an usual way as $\xi=-\frac{\Delta.n}{2P.n}$.

We  can then express the momenta of the particles through their  
Sudakov decomposition and, keeping the first-order corrections in the masses and $\Delta_T^2$, we have:
\eqs{\label{eq:decomp_momenta}
p_1&= (1+\xi) p + \frac{M^2}{1+\xi}n,\\
q&\simeq- 2 \xi \Big(1+ \frac{(\Delta_T^2-M^2)}{Q^2}\Big)  p + \frac{Q^2}{2\xi\Big(1+ \textstyle \frac{(\Delta_T^2-M^2)}{Q^2}\Big)} n,\\
p_\pi&= (1-\xi) p +\frac{m_\pi^2-\Delta_T^2}{1-\xi}n+ \Delta_T,\nn}
\eqs{\label{eq:decomp_momentb}
p_2&\simeq- 2 \xi \frac{(\Delta_T^2-M^2)}{Q^2} p\;+ \\&\Big[\frac{Q^2}{2\xi\Big(1+ \textstyle \frac{(\Delta_T^2-M^2)}{Q^2}\Big)} -\frac{m_\pi^2-\Delta_T^2}{1-\xi}+ \frac{M^2}{1+\xi}\Big]n - \Delta_T,\\
\Delta&= - 2 \xi p +\Big[\frac{m_\pi^2-\Delta_T^2}{1-\xi}- \frac{M^2}{1+\xi}\Big]n
+ \Delta_T.}

The polarisation vectors of the virtual photon are chosen to be (in the $P'\pi^0$ center-of-mass frame):
\eqs{\label{eq:pol_gamma}
&\ep_L(q)\simeq\frac{2\xi(Q^2+\Delta_T^2-M^2)}{Q^3}p +\frac{Q^3}{2\xi(Q^2+\Delta_T^2-M^2)}n ,\\&\ep_{T_1}(q)=\ep_x, ~~\ep_{T_2}(q)=\ep_y.
}

We also have
\eqs{
W^2=(p_1+q)^2\simeq \frac{(1-\xi)Q^2}{2\xi}- (\Delta_T^2-M^2),
}
where we have kept the leading term in $Q^2$ and the next-to-leading
one which does not vanish in the limit $\xi\to 1$.
This provides us with the following relation between $\xi$ and $W^2$
\eqs{
\xi\simeq\frac{Q^2}{Q^2+2(W^2+\Delta_T^2-M^2)},
}
which reduces to the usual one, $\xi\simeq\frac{Q^2}{Q^2+2W^2}$, when 
$M^2$ and $\Delta_T^2$ can be neglected compared to $W^2$ (which is not
the case in the $\xi\to 1$ limit).
Furthermore, we have the exact relation
\eqs{
x_B=\frac{Q^2}{2p_1.q}=\frac{Q^2}{W^2+Q^2-M^2},
}
which gives 
\eqs{
\xi \simeq \frac{x_B}{2-x_B}.
}

Finally, we have (neglecting the pion mass):
\eqs{
\Delta_T^2=\frac{1-\xi}{1+\xi} u-2 M^2 \frac{1-\xi}{(1+\xi)^2}\xi.
}

In Ref.~\cite{TDApiproton}, we have defined the leading-twist proton to pion
$P \to \pi$ transition distribution amplitudes from the Fourier transform
of the matrix element

\eqs{\label{eq:mat_el_p-pi}
   \langle \pi|\, \epsilon^{ijk} {q}^{i'}_{\alpha}(z_{1}n)\, 
[z_1;z_0]_{i',i}\,{q}^{j'}_{\beta}(z_{2} n)\, [z_2;z_0]_{j'j}\,
   {q}^{k'}_{\gamma}(z_{3} n)\\ \times[z_3;z_0]_{k'k} \,|P \rangle.
}

The brackets $[z_i;z_0]$ in \ce{eq:mat_el_p-pi} account for the insertion of a 
path-ordered gluonic exponential 
along the straight line connecting an arbitrary initial point $z_0 n$ and a final one $z_i n$: 
\eqs{
   [z_i;z_0] \equiv {\rm P\ exp\,}
   \biggl[ ig\!\!\!\int_0^1\!\!\!dt\, (z_i-z_0)n_\mu A^\mu(n[tz_i+(1-t)z_0])\biggr]
}
 which provide the QCD-gauge invariance for such non-local operator and
equal unity in a light-like (axial) gauge.

\begin{widetext}
The leading-twist TDAs for the $p \to \pi^0$ transition, $ V^{p\pi^0}_{i}\!\!(x_i,\xi, \Delta^2)$, 
$A^{p\pi^0}_{i}\!\!(x_i,\xi, \Delta^2)$ and 
$T^{p\pi^0}_{i}\!\!(x_i,\xi, \Delta^2)$  are defined here\footnote{The present definitions differ from those 
of~\cite{TDApiproton} by constant multiplicative factors and by the definition of $\sigma^{\mu\nu}$.}  
as\footnote{In the following,
we shall use the notation $\displaystyle {\cal F}\equiv (p.n)^3\int^{\infty}_{-\infty} \Pi_j dz_j/(2\pi)^3 e^{i\Sigma_k x_k z_k p.n}$.} :
\bea{TDA}
 && 4 {\cal F}\Big(\langle     \pi^0(p_\pi)|\, \epsilon^{ijk}u^{i}_{\alpha}(z_1 n) 
u^{j}_{\beta}(z_2 n)d^{k}_{\gamma}(z_3 n)
\,|P(p_1,s_1) \rangle \Big)  
 =   
\\ \nonumber
&&
i\frac{f_N}{f_\pi}\Big[ V^{p\pi^0}_{1} (\ks p C)_{\alpha\beta}(N^+)_{\gamma}+A^{p\pi^0}_{1} (\ks p\gamma^5 C)_{\alpha\beta}(\gamma^5 N^+)_{\gamma} +
T^{p\pi^0}_{1} (\sigma_{p\mu} C)_{\alpha\beta}(\gamma^\mu N^+)_{\gamma} 
\nonumber \\
&& + M^{-1}V^{p\pi^0}_{2} 
 (\ks p C)_{\alpha\beta}(\ks \Delta\!_T N^+)_{\gamma} +M^{-1}
A^{p\pi^0}_{2}(\ks p \gamma^5 C)_{\alpha\beta}(\gamma^5\ks \Delta\!_T N^+)_{\gamma}
+ M^{-1}T^{p\pi^0}_{2} ( \sigma_{p\Delta_T} C)_{\alpha\beta}(N^+)_{\gamma}
\nonumber \\
&&+  M^{-1}T^{p\pi^0}_{3} ( \sigma_{p\mu} C)_{\alpha\beta}(\sigma^{\mu\Delta_T}
 N^+)_{\gamma} + M^{-2}T^{p\pi^0}_{4} (\sigma_{p \Delta_T} C)_{\alpha\beta}
(\ks \Delta\!_T N^+)_{\gamma}\;\Big], \nonumber
\eea

~\\~\\
where $\sigma^{\mu\nu}= 1/2[\gamma^\mu, \gamma^\nu]$ with $\sigma^{p \mu} = p_\nu \sigma^{\nu\mu}$,..., $C$ is the charge 
conjugation matrix 
and $N^+$ is the large component of the nucleon spinor 
($N=(\ks n \ks p + \ks p \ks n) N = N^-+N^+$
with $N^+\sim \sqrt{p_1^+}$ and $N^-\sim \sqrt{1/p_1^+}$).
 $f_\pi$ is the pion decay constant ($f_\pi = 131$ MeV) and $f_N$ has been estimated through 
QCD sum rules to be of
order $5.2\cdot 10^{-3}$ GeV$^2$~\cite{CZ}. All the TDAs $V_i$, $A_i$ and $T_i$ are
dimensionless. Note that the first three 
terms in (\ref{TDA}) are the only ones surviving the limit $\Delta_T \to 0$.

\end{widetext}

  \section{The soft-pion limit}\label{sec:softpionlimit}

We now  derive the general limit
of these three contributing TDAs at $\Delta_T=0$ in the soft-pion limit, when $\xi$ gets close
to 1 (see also~\cite{Lansberg:2007se}). In that limit, the soft-meson theorem~\cite{AD} 
derived from current algebra
apply~\cite{Pobylitsa:2001cz}, which allow us to express these 3 TDAs in terms  of the 3 Distribution
Amplitudes (DAs) of the corresponding baryon. In the case of the proton DA~\cite{CZ}, 
$V^p(x_i)$, $A^p(x_i)$, $T^p(x_i)$ are defined such as
\eqs{\label{eq:DA}
4{\cal F}_{DA}\Big(\langle 0|\epsilon_{ijk}u^i_\alpha(z_1 n)&u^j_\beta(z_2 n)d^k_\gamma(z_3 n)|P(p,s)\rangle\Big)
  = f_N \times \\\Big[
&V^p(x_{i}) ( \ks p C)_{\alpha \beta} (\gamma^5 N^+)_\gamma 
\\ +&A^p(x_{i}) (\ks p \gamma^5 C)_{\alpha \beta} N^+_\gamma
\\ +&T^p(x_{i}) (\sigma_{p  \mu}\,C)_{\alpha \beta} 
(\gamma^\mu \gamma^5 N^+)_\gamma 
\Big],
}
with $ {\cal F}_{DA}\equiv (p.n)^3\int^{\infty}_{-\infty} \Pi_j \frac{dz_j}{(2\pi)^3} e^{i\Sigma_k x_k z_k p.n}$ where $p$ is here the proton momentum.

Inspired by~\cite{Pobylitsa:2001cz}, which considered the related case of the 
distribution amplitude of the proton-meson system, we use the soft pion 
theorems~\cite{AD} to write:

\begin{eqnarray}
&&\langle \pi^a(p_\pi)  |{\cal O}| N(p_1,s_1)\rangle = -\frac{i}{f_\pi} \langle 0  | [ Q^a_5,  {\cal O}]  | N(p_1,s_1) \rangle \\ \nonumber
&&\!\!\!\! +\frac{ig_A} {4f_\pi p_1.p_\pi }
\sum_{s'_1} \bar N (p_1,s'_1)\ks p_\pi \gamma_5 \tau^a N(p_1,s_1) \langle 0 |{\cal O}| N(p_1,s'_1)\rangle 
\label{eq:soft-theorem}
\end{eqnarray}

The second term, which takes care of the nucleon pole term, does not contribute at threshold and
will not be considered in the following.

For the transition $p\to \pi^0$, $Q^a_5=Q^3_5$ and ${\cal O}=u_\alpha u_\beta d_\gamma$. 
Since the commutator of the chiral charge $Q_{5}$ with the quark field $\psi$ ($\tau^a $ being the Pauli 
matrix)
\begin{equation}
[Q_{5}^a, \psi] = - \frac{\tau^a}{2} \gamma^5 \psi\;,
\label{eq:chiral-trans}
\end{equation}
the first term in the rhs of \ce{eq:soft-theorem} gives three terms from 
$(\ga u)_\alpha u_\beta d_\gamma$, $u_\alpha (\ga u)_\beta d_\gamma$ and $u_\alpha u_\beta (\ga d)_\gamma$.
The corresponding multiplication by $\ga$ (or $(\ga)^T$ when it acts on the index $\beta$) on the 
vector and axial structures of the DA (\ce{eq:DA}) gives two terms which cancel and the third one,
which remains, is the same as the one for the TDA up to the modification that on the DA decomposition,
$p$ is the proton momentum, whereas for the TDA one, $p$ is the light-cone projection of $P$, $\ie$ half
the proton momentum if $\xi=1$. This introduces a factor $2\xi$ in the relations
between the 2 DAs $A^p$ and $V^p$ and the 2 TDAs $V^{p\pi^0}_{1}$ and $A^{p\pi^0}_{1}$, which is 
canceled though by the factor one half in \ce{eq:chiral-trans}. 

To what concerns the tensorial structure multiplying $T^p$,  the three 
terms are identical at leading-twist accuracy and corresponds to the structure multiplying
$T^{p\pi^0}_{1}$, this gives a factor 3. Finally, an extra factor $(4\xi)^{-1}$ appears when one goes to
the momentum space~\cite{Lansberg:2007se}. We eventually have at $\Delta_T=0$:

\bea{eq:softp}
&&V^{p\pi^0}_1(x_1,x_2,x_3,\xi,\Delta^2) =  \frac{V^p}{4 \xi}  \Big(\frac{x_1}{2\xi},\frac{x_2}{2\xi},\frac{x_3}{2\xi} \Big), 
\nonumber \\
&&A^{p\pi^0}_1(x_1,x_2,x_3,\xi,\Delta^2) = \frac{A^p}{4 \xi}  \Big(\frac{x_1}{2\xi},\frac{x_2}{2\xi},\frac{x_3}{2\xi} \Big), 
 \\
&&T^{p\pi^0}_1(x_1,x_2,x_3,\xi,\Delta^2) = \frac{3 T^p}{4 \xi}  \Big(\frac{x_1}{2\xi},\frac{x_2}{2\xi},\frac{x_3}{2\xi} \Big). 
 \nonumber  
\eea

Note the factor $\frac{1}{2\xi}$ in the argument of the DA in~\ce{eq:softp}. We refer to~\cite{Lansberg:2007se} for a complete discussion. Indeed, for the TDAs, 
the $x_i$ are defined with respect to $p$ 
( see \eg~$ {\cal F}\equiv (p.n)^3\int^{\infty}_{-\infty} \Pi_j \frac{dz_j}{(2\pi)^3} e^{i\Sigma_k x_k z_k p.n}$) which tends to $\frac{p_1}{2}$ when $\xi \to 1$. Therefore, they
vary within the interval $[-2,2]$, whereas for the DAs, the momentum fractions are defined with
respect to the proton momentum $p_1$ and vary between 0 and 1.

Our results are comparable to the ones for the proton-pion DAs obtained in~\cite{Braun:2006td}.
Finally, it is essential to note that these limiting values are not zero, unlike for some GPDs. 
Hence, we find it reasonable to conjecture that these expressions give the right order of magnitude 
of the TDAs for quite large values of $\xi$ (say $\xi \geq 0.5 $) in a first estimate of cross sections.

 \section{Hard-amplitude calculation}

At leading order in $\alpha_s$, the amplitude ${\cal M}^\lambda_{s_1s_2}$  for 
$\gamma^\star(q,\lambda) P(p_1,s_1) \to P'(p_2,s_2) \pi^0(p_\pi)$ reads

\eqs{\label{eq:ampl-bEPM1}
&{\cal M}^\lambda_{s_1s_2}=-i 
\frac{(4 \pi \alpha_s)^2 \sqrt{4 \pi \alpha_{em}} f_{N}^2}{ 54 f_{\pi}Q^4}\times\\&\Big[ 
\underbrace{ \bar u(p_2,s_2) \ks \ep(\lambda) \gamma^5 u(p_1,s_1)}_{{\cal S}^\lambda_{s_1s_2}}\\&
~~~~~~~~~~~~~~~~~~~~~~
\times\underbrace{\int\limits^{1+\xi}_{-1+\xi} \! \! \! d^3x \int\limits_0^1 \! \!d^3y
\Bigg(2\sum\limits_{\alpha=1}^{7} T_{\alpha}+
\sum\limits_{\alpha=8}^{14} T_{\alpha}\Bigg)}_{{\cal I}}\\&-
\underbrace{\ep(\lambda)_\mu \Delta_{T,\nu} \bar u(p_2,s_2) (\sigma^{\mu \nu}+ g^{\mu\nu}) \gamma^5 u(p_1,s_1) }_{{\cal S'}^\lambda_{s_1s_2}}\\&
~~~~~~~~~~~~~~~~~~~
\times\underbrace{\int\limits^{1+\xi}_{-1+\xi} \! \! \! d^3x \int\limits_0^1 \! \!d^3y
\Bigg(2\sum\limits_{\alpha=1}^{7} T'_{\alpha}+
\sum\limits_{\alpha=8}^{14} T'_{\alpha}\Bigg)}_{{\cal I'}}\Big]
,}
where the coefficient $T_{\alpha}$ and $T'_{\alpha}$$(\alpha=1,...,14)$ are functions of $x_{i}$,$y_{j}$,$\xi$ and $\Delta^2$ and  are given in \ct{tab:coeff_funct}. In general, 
we have 21 diagrams: we have not drawn  7 others which differ only to the 7 first ones by
a permutation between the $u$-quark lines $1$ and $2$. The symmetry of the integration domain 
and of the DAs and TDAs with respect to the changes $x_1 \leftrightarrow x_2$ and 
$y_1 \leftrightarrow y_2$ therefore tells us that they give the same contributions as the 7 first
diagrams. They are accounted for by a factor 2 in the last equation.

The integrals in~\ce{eq:ampl-bEPM1} are understood with two delta-functions insuring momentum conservation:
 \eqs{&\int d^3x \equiv \int dx_1dx_2dx_3\delta(2\xi -x_1-x_2-x_3)} and 
\eqs{&\int d^3y \equiv \int dy_1dy_2dy_3\delta(1-y_1-y_2-y_3).}

The expression in \ce{eq:ampl-bEPM1} is to be compared with the leading-twist amplitude for the 
baryonic-form factor~\cite{CZ}
\eqs{\label{eq:ampl-FF} &{\cal M}^\lambda \propto  -i\bar u(p_2) \ks \ep(\lambda) u(p_1)\frac{\alpha_s^2 
f_{N}^2}{Q^4}  
\\&\int\limits_{0}^{1} d^3x \int\limits_0^1 d^3y
\Bigg(2\sum\limits_{\alpha=1}^{7} T^p_{\alpha}(x_{i},y_{j},\xi,t)+
\sum\limits_{\alpha=8}^{14} T^p_{\alpha}(x_{i},y_{j},\xi,t)\Bigg)
.}

The factors $T^p_{\alpha}$ are very similar to the $T_{\alpha}$ obtained here.

\begin{widetext}
\center
\begin{longtable}{cccc}
KILLED & LINE!!!! \kill
\caption{14 of the 21 diagrams contributing to the hard-scattering amplitude with
their associated coefficient $T_\alpha$ and $T'_\alpha$. The seven first ones with $u$-quark
lines inverted are not drawn. The crosses represent the virtual-photon vertex.}
\\\hline\hline& &  \\
$\alpha$ & & $T_\alpha$& $T'_\alpha$ \\
\\\hline& &  \\
\parbox{.15cm}{1\\ ~\\ ~\\}&
\includegraphics[height=0.9001cm,clip=true]{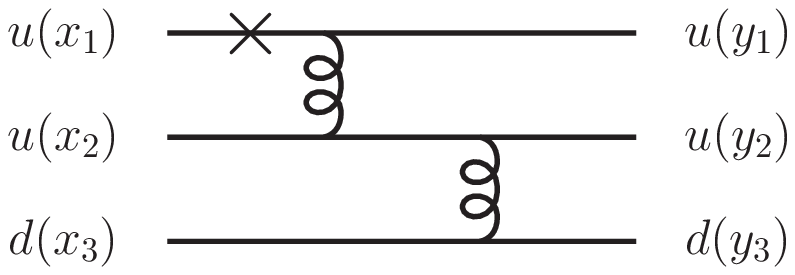} &  \parbox{7.21cm}{\vspace*{-.7cm}
$   \frac{-Q_u(2\xi)^2[(V_1^{p\pi^0}-A_1^{p\pi^0})(V^p-A^p)+4T^{p\pi^0}_1 T^p+2\frac{\Delta^2_T}{M^2}T^{p\pi^0}_4 T^p]}
{(2\xi-x_{1}-i\epsilon)^2(x_{3}-i\epsilon)(1-y_{1})^2y_{3}}$ \\ ~ \\ }&  \parbox{6.76cm}{\vspace*{-.7cm}
$  \frac{-Q_u(2\xi)^2[(V_2^{p\pi^0}-A_2^{p\pi^0})(V^p-A^p)+2(T^{p\pi^0}_2+T^{p\pi^0}_3) T^p]}
{(2\xi-x_{1}-i\epsilon)^2(x_{3}-i\epsilon)(1-y_{1})^2y_{3}}$ \\ ~ \\ }\\ 
\parbox{.15cm}{2\\ ~\\ ~\\}&
\includegraphics[height=0.9001cm,clip=true]{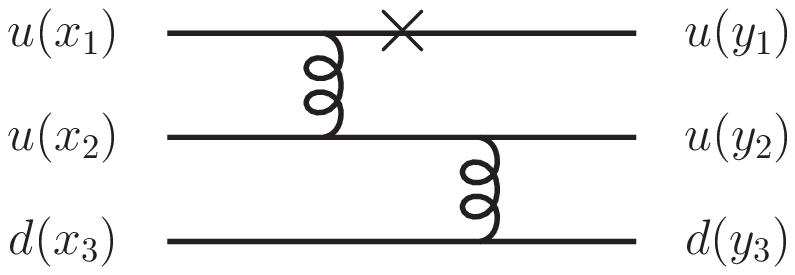} & \parbox{7.21cm}{\vspace*{-.7cm}
$ 0 $\\ ~ \\ }& \parbox{6.76cm}{\vspace*{-.7cm}
$ 0 $\\ ~ \\ }\\ 
\parbox{.15cm}{3\\ ~\\ ~\\}&
\includegraphics[height=0.9001cm,clip=true]{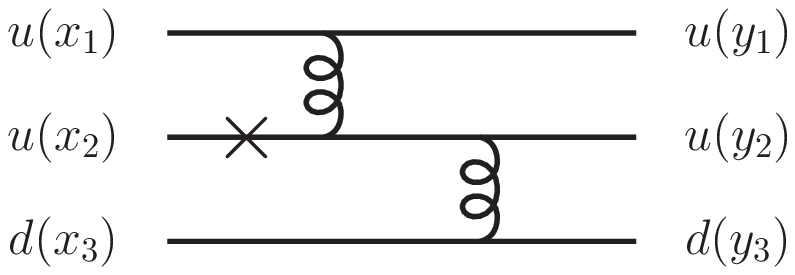} & \parbox{7.21cm}{\vspace*{-.7cm}
$ \frac{Q_u(2\xi)^2[4T^{p\pi^0}_1 T^p+2\frac{\Delta^2_T}{M^2}T^{p\pi^0}_4 T^p]}
{(x_{1}-i\epsilon) (2\xi-x_{2}-i\epsilon)(x_{3}-i\epsilon)y_{1}(1-y_{1})y_{3}}$\\ ~ \\ }&  \parbox{6.76cm}{\vspace*{-.7cm}
$ \frac{Q_u(2\xi)^2[2(T^{p\pi^0}_2+T^{p\pi^0}_3) T^p]}
{(x_{1}-i\epsilon) (2\xi-x_{2}-i\epsilon)(x_{3}-i\epsilon)y_{1}(1-y_{1})y_{3}}$ \\ ~ \\ }\\ 
\parbox{.15cm}{4\\ ~\\ ~\\}&
\includegraphics[height=0.9001cm,clip=true]{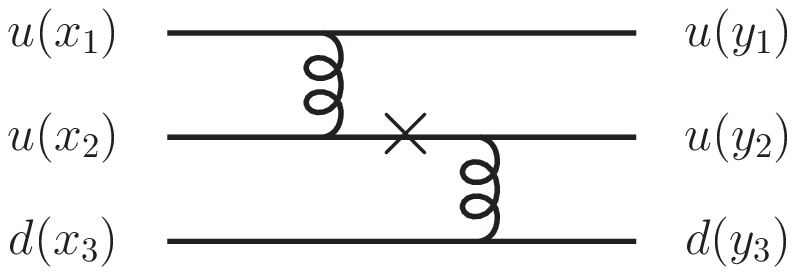} & \parbox{7.21cm}{\vspace*{-.7cm}
$ \frac{-Q_u(2\xi)^2[(V_1^{p\pi^0}-A_1^{p\pi^0})(V^p-A^p)]
}{(x_{1}-i\epsilon)
(2\xi-x_{3}-i\epsilon)(x_{3}-i\epsilon)y_{1}(1-y_{1})y_{3}}$ \\ ~ \\ }&  \parbox{6.76cm}{\vspace*{-.7cm}
$  \frac{-Q_u(2\xi)^2[(V_2^{p\pi^0}-A_2^{p\pi^0})(V^p-A^p)]}
{(x_{1}-i\epsilon)
(2\xi-x_{3}-i\epsilon)(x_{3}-i\epsilon)y_{1}(1-y_{1})y_{3}}$ \\ ~ \\ }\\ 
\parbox{.15cm}{5\\ ~\\ ~\\}&
\includegraphics[height=0.9001cm,clip=true]{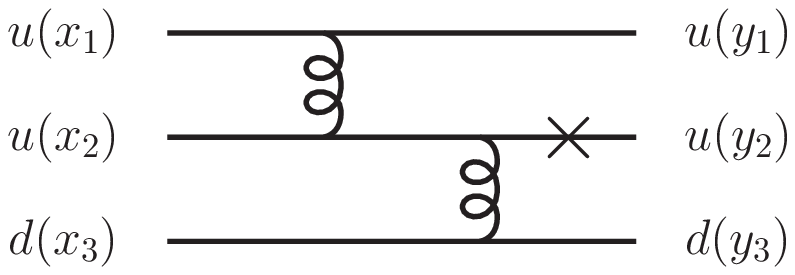} & \parbox{7.21cm}{\vspace*{-.7cm}
$ \frac{Q_u(2\xi)^2[(V_1^{p\pi^0}+A_1^{p\pi^0})(V^p+A^p)]}
{(x_{1}-i\epsilon)(2\xi-x_{3}-i\epsilon)(x_{3}-i\epsilon)y_{1}(1-y_{2})y_{3}}$ \\ ~ \\ }&  \parbox{6.76cm}{\vspace*{-.7cm}
$ \frac{Q_u(2\xi)^2[(V_2^{p\pi^0}+A_2^{p\pi^0})(V^p+A^p)]}
{(x_{1}-i\epsilon)(2\xi-x_{3}-i\epsilon)(x_{3}-i\epsilon)y_{1}(1-y_{2})y_{3}}$ \\ ~ \\ }\\ 
\parbox{.15cm}{6\\ ~\\ ~\\}&
\includegraphics[height=0.9001cm,clip=true]{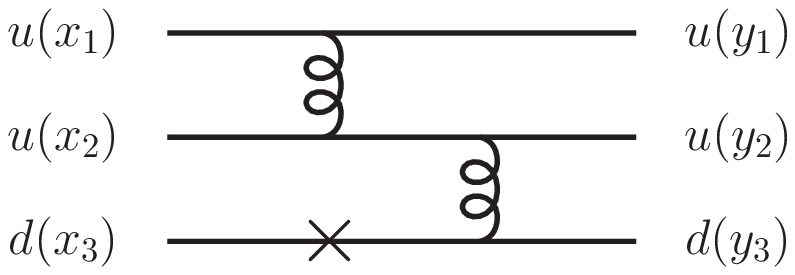} & \parbox{7.21cm}{\vspace*{-.7cm}
$ 0$\\ ~ \\ }& \parbox{6.76cm}{\vspace*{-.7cm}
$ 0 $\\ ~ \\ }\\ 
\parbox{.15cm}{7\\ ~\\ ~\\}&
\includegraphics[height=0.9001cm,clip=true]{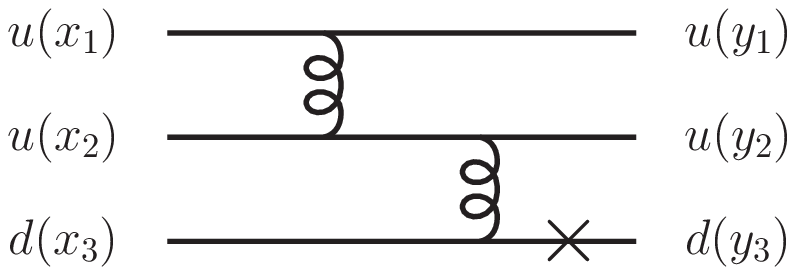} & \parbox{7.21cm}{\vspace*{-.7cm}
$ \frac{-Q_d(2\xi)^2[2(V_1^{p\pi^0}V^p+A_1^{p\pi^0}A^p)]
}{(x_{1}-i\epsilon)(2\xi-x_{3}-i\epsilon)^2y_{1}(1-y_3)^2} $\\ ~ \\ }&  \parbox{6.76cm}{\vspace*{-.7cm}
$ \frac{-Q_d(2\xi)^2[2(V_2^{p\pi^0}V^p+A_2^{p\pi^0}A^p)]}
{(x_{1}-i\epsilon)(2\xi-x_{3}-i\epsilon)^2y_{1}(1-y_3)^2}$ \\ ~ \\ }\\ 
\parbox{.15cm}{8\\ ~\\ ~\\}&
\includegraphics[height=0.9001cm,clip=true]{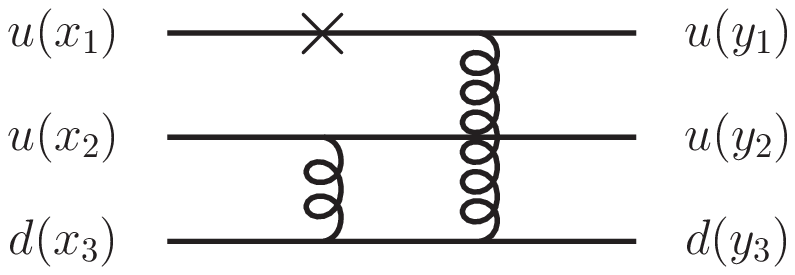} & \parbox{7.21cm}{\vspace*{-.7cm}
$ 0$\\ ~ \\ }& \parbox{6.76cm}{\vspace*{-.7cm}
$ 0 $\\ ~ \\ }\\
\parbox{.15cm}{9\\ ~\\ ~\\}&
\includegraphics[height=0.9001cm,clip=true]{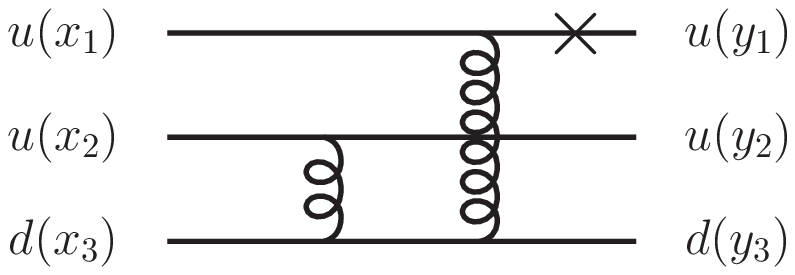} & \parbox{7.21cm}{\vspace*{-.7cm}
$ \frac{-Q_u(2\xi)^2[(V_1^{p\pi^0}-A_1^{p\pi^0})(V^p-A^p)+ 4T^{p\pi^0}_1 T^p+2\frac{\Delta^2_T}{M^2}T^{p\pi^0}_4 T^p]}
{(2\xi-x_{1}-i\epsilon)^2(x_{2}-i\epsilon)(1-y_{1})^2y_{2}} $\\ ~ \\ }&  \parbox{6.76cm}{\vspace*{-.7cm}
$ \frac{-Q_u(2\xi)^2[(V_2^{p\pi^0}-A_2^{p\pi^0})(V^p-A^p)+2(T^{p\pi^0}_2+T^{p\pi^0}_3) T^p]}
{(2\xi-x_{1}-i\epsilon)^2(x_{2}-i\epsilon)(1-y_{1})^2y_{2}}$ \\ ~ \\ }\\ 
\parbox{.15cm}{10\\ ~\\ ~\\}&
\includegraphics[height=0.9001cm,clip=true]{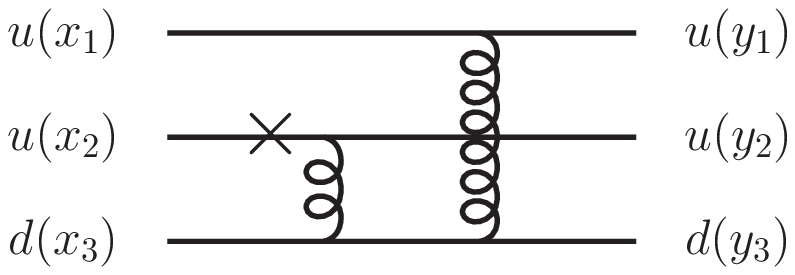} & \parbox{7.21cm}{\vspace*{-.7cm}
$ \frac{-Q_u(2\xi)^2[(V_1^{p\pi^0}+A_1^{p\pi^0})(V^p+A^p)+ 4T^{p\pi^0}_1 T^p+2\frac{\Delta^2_T}{M^2}T^{p\pi^0}_4 T^p]}
{(x_{1}-i\epsilon)(2\xi-x_{2}-i\epsilon)^2y_{1}(1-y_{2})^2}$\\ ~ \\ }&  \parbox{6.76cm}{\vspace*{-.7cm}
$ \frac{-Q_u(2\xi)^2[(V_2^{p\pi^0}+A_2^{p\pi^0})(V^p+A^p)+2(T^{p\pi^0}_2+T^{p\pi^0}_3) T^p]}
{(x_{1}-i\epsilon)(2\xi-x_{2}-i\epsilon)^2y_{1}(1-y_{2})^2}$ \\ ~ \\ }\\ 
\parbox{.15cm}{11\\ ~\\ ~\\}&
\includegraphics[height=0.9001cm,clip=true]{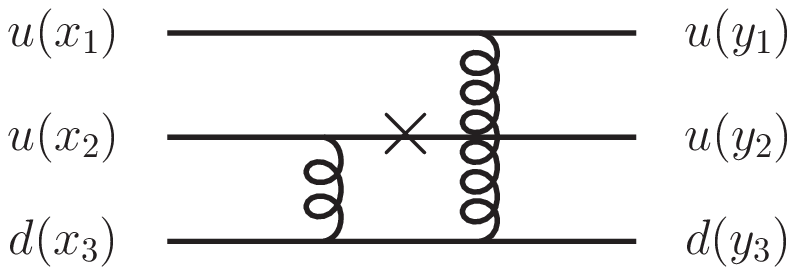} & \parbox{7.21cm}{\vspace*{-.7cm}
$ 0$\\ ~ \\ }& \parbox{6.76cm}{\vspace*{-.7cm}
$ 0 $\\ ~ \\ }\\ 
\parbox{.15cm}{12\\ ~\\~\\ }&
\includegraphics[height=0.9001cm,clip=true]{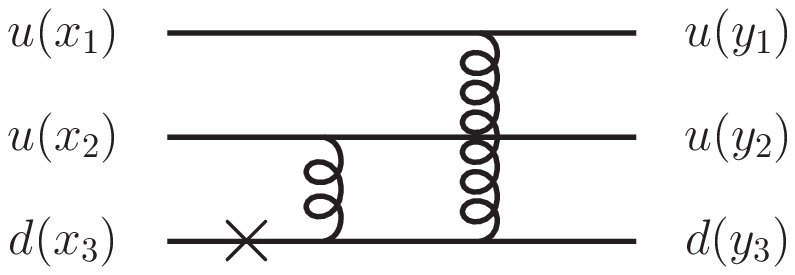} & \parbox{7.21cm}{\vspace*{-.7cm}
$ \frac{Q_d(2\xi)^2[(V_1^{p\pi^0}+A_1^{p\pi^0})(V^p+A^p)]}
{(x_{1}-i\epsilon)(x_{2}-i\epsilon)(2\xi-x_{3}-i\epsilon)y_{1}(1-y_{2})y_{2}}$\\ ~ \\ }&  \parbox{6.76cm}{\vspace*{-.7cm}
$ \frac{Q_d(2\xi)^2[(V_2^{p\pi^0}+A_2^{p\pi^0})(V^p+A^p)]}
{(x_{1}-i\epsilon)(x_{2}-i\epsilon)(2\xi-x_{3}-i\epsilon)y_{1}(1-y_{2})y_{2}}$ \\ ~ \\ }\\ 
\parbox{.15cm}{13\\ ~\\ ~\\}&
\includegraphics[height=0.9001cm,clip=true]{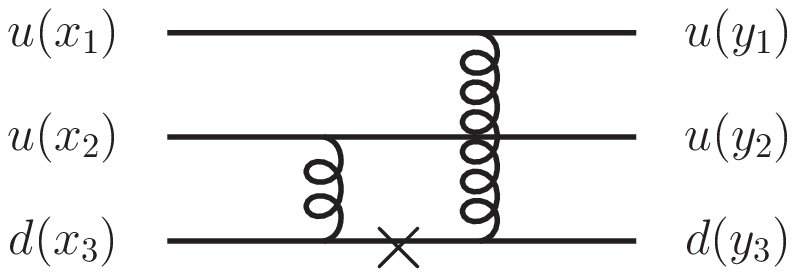} & \parbox{7.21cm}{\vspace*{-.7cm}
$ \frac{ -Q_d(2\xi)^2[4 T^{p\pi^0}_1 T^p+2\frac{\Delta^2_T}{M^2}T^{p\pi^0}_4 T^p]}
{(x_{1}-i\epsilon)(2\xi-x_{1}-i\epsilon)(x_{2}-i\epsilon)y_{1}(1-y_{2})y_{2}}$\\ ~ \\ }&  \parbox{6.76cm}{\vspace*{-.7cm}
$ \frac{-Q_d(2\xi)^2[2 (T^{p\pi^0}_2+T^{p\pi^0}_3)T^p]}
{(x_{1}-i\epsilon)(2\xi-x_{1}-i\epsilon)(x_{2}-i\epsilon)y_{1}(1-y_{2})y_{2}}$ \\ ~ \\ }\\ 
\parbox{.15cm}{14\\ ~\\ ~\\}&
\includegraphics[height=0.9001cm,clip=true]{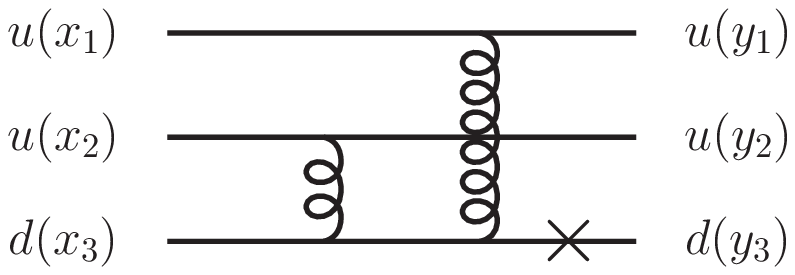} & \parbox{7.21cm}{\vspace*{-.7cm}
$ \frac{Q_d(2\xi)^2[(V_1^{p\pi^0}-A_1^{p\pi^0})(V^p-A^p)]}
{(x_{1}-i\epsilon)(2\xi-x_{1}-i\epsilon)(x_{2}-i\epsilon)y_{1}y_{2}(1-y_{3})}$\\ ~ \\ }&  \parbox{6.76cm}{\vspace*{-.7cm}
$ \frac{Q_d(2\xi)^2[(V_2^{p\pi^0}-A_2^{p\pi^0})(V^p-A^p)]}
{(x_{1}-i\epsilon)(2\xi-x_{1}-i\epsilon)(x_{2}-i\epsilon)y_{1}y_{2}(1-y_{3})}$ \\ ~ \\ }\\ 
\label{tab:coeff_funct}
\end{longtable}

\end{widetext}

\section{Cross-section estimate for unpolarised protons}

When $\xi$ is large, the ERBL region covers most of the integration domain. This corresponds to the 
emission of three quarks of positive light-cone-momentum fraction off the target proton. Therefore 
it is legitimate 
to approximate the cross section only from the ERBL region.  As a consequence, the integration on 
the momentum fractions contained in the TDAs
between $-1+\xi$ and ${1+\xi}$ (see~\ce{eq:ampl-bEPM1}) can be converted into one between 0 and 1 by 
a change of variable and can be carried out straightforwardly.

On the other hand, we have at our disposal a reasonable estimation of the 
TDAs $V_1^{p\pi^0}$, $A_1^{p\pi^0}$ and $T_1^{p\pi^0}$ in the large-$\xi$ region and for 
vanishing $\Delta_T$ thanks to the soft pion limit (see section~\ref{sec:softpionlimit}).
As a consequence, we have all 
the tools needed for a first evaluation
of the unpolarised cross section  for $\gamma^\star P \to P' \pi^0$ for large  $\xi$ and 
when $\Delta^2_T$ is vanishing.

The differential cross section for  unpolarised protons in the proton-pion center-of-mass frame is 
calculated as usual from the 
averaged-squared amplitudes, ${|{\cal  M}_i|}^2$ ($i=T,L,TT,LT$):
\eqs{
\frac{d \sigma_i}{d\Omega^*_{\pi}}=& \frac{1}{64 \pi^2 W^2} \frac{\sqrt{(p_\pi.p_2)^2-m_\pi^2 M^2)}}
{\sqrt{(p_1.q)^2+M^2 Q^2)}}{|{\cal  M}_i|}^2\\
\simeq&\frac{W^2-M^2}{64 \pi^2 W^2 \sqrt{(W^2+M^2+Q^2)^2-4W^2 M^2}}{|{\cal  M}_i|}^2.}
The ${|{\cal  M}_i|}^2$ are obtained from squaring and summing (resp. averaging) ${\cal M}^\lambda_{s_1s_2}$ over the final (resp. initial) proton helicities with given appropriate combinations of 
the photon helicity $\lambda$~\cite{Mulders:1990xw}. The expression of ${\cal M}^\lambda_{s_1s_2}$
are obtained from~\ce{eq:ampl-bEPM1}.

For vanishing $\Delta^2_T$, the spinorial structure ${\cal S}^\lambda_{s_1s_2}$ 
\eqs{\bar u(p_2,s_2) \ks \ep(\lambda) \gamma^5 u(p_1,s_1)}
only survives. To obtain ${|{\cal  M}_T|}^2$, we square the latter and 
sum over the proton helicities and the transverse polarisations
of the photon, it gives a factor  $\frac{2(1+\xi) Q^2}{\xi}$. On the other hand, 
${|{\cal  M}_L|}^2$, vanishes at the leading-twist accuracy, 
as in the nucleon-form-factor case. The same is true for $|{{\cal  M}_{LT}}|^2$ and
${|{\cal  M}_{TT}|}^2$ since the  $x$ and $y$ direction are not distinguishable
when  $\Delta^2_T$ is vanishing.

Contrariwise, if we wanted to consider the spinorial structure ${\cal S'}^\lambda_{s_1s_2}$ -- arising when 
 $\Delta^2_T\neq 0$ --
\eqs{\ep(\lambda)_\mu \Delta_{T,\nu} \bar u(p_2,s_2) (\sigma^{\mu \nu}+ g^{\mu\nu}) \gamma^5 u(p_1,s_1)\; ,}
$|{{\cal  M}_{TT}}|^2$, and thus $\frac{d \sigma_{TT}}{d\Omega^*_{\pi}}$, would not be zero and 
the cross section would show a $\cos 2 \varphi$ dependence.

The remaining
part still to be considered is now entirely contained in the factor $\cal I$ of~\ce{eq:ampl-bEPM1} for which
we need to choose parametrisations for the DAs and the TDAs.
For the sake of coherence, we shall choose the same parametrisation for both.
 Since the asymptotic limit~\cite{Lepage:1980fj} $V^p(x_i)=T^p(x_i)=\varphi_{as}= 120 x_1 x_2 x_3$ 
and $A(x_i)=0$
is known to give a vanishing proton-form factor and the wrong sign
to the neutron one, we shall not use it.

Note, however, that the isospin relations between the TDAs $V^{p\pi^0}_{1}$, $A^{p\pi^0}_{1}$ and
$T^{p\pi^0}_{1}$ differ from those between the DAs; the factor 3 in \ce{eq:softp} clearly illustrates this
fact. Therefore, whereas the asymptotic limit choices give a vanishing proton form factor due to 
a full cancellation between the 14-diagram contributions, the resulting expression will not vanish here
even for the asymptotic DAs and TDAs derived in the soft limit. 

Yet, we shall rather consider the more reasonable choices of 
V.~L.~Chernyak and A.~R.~Zhitnitsky~\cite{CZ} 
(noted CZ) based on an analysis of QCD sum rules. 

Therefore, we take for the DAs:
\eqs{\label{eq:DACZ}
V^p(x_i)&=\varphi_{as} [11.35 (x_1^2+x_2^2)+8.82 x_3^2-1.68 x_3 -2.94],\\
A^p(x_i)&=\varphi_{as} [6.72 (x_2^2-x_1^2)],\\
T^p(x_i)&=\varphi_{as} [13.44 (x_1^2+x_2^2)+4.62 x_3^2+0.84 x_3 -3.78],\\
}
and for the limiting value of the TDAs
\eqs{\label{eq:TDACZ}
V^{p\pi^0}\!\!&=\frac{\varphi_{as}}{2^5\xi^4}[\frac{11.35}{(2\xi)^2} (x_1^2+x_2^2)+\frac{8.82}{(2\xi)^2} x_3^2-\frac{1.68}{2\xi} x_3 -2.94], \\
A^{p\pi^0}\!\!&=\frac{\varphi_{as}}{2^5\xi^4}[\frac{6.72}{(2\xi)^2} (x_2^2-x_1^2)], \\
T^{p\pi^0}\!\!&=\frac{3\varphi_{as}}{2^5\xi^4} [\frac{13.44}{(2\xi)^2} (x_1^2+x_2^2)+\frac{4.62}{(2\xi)^2} x_3^2+\frac{0.84}{2\xi} x_3 -3.78].\\
}

With this choice, we get the following analytic result valid at large values of $\xi$:
\eqs{
\frac{d \sigma}{d\Omega^*_{\pi}}&\Big|_{\theta^*_\pi=\pi}=\frac{0.389\alpha_{em} 
\alpha_s^4 f_N^4 \pi^3(W^2-M^2)}{f^2_\pi W^2 \sqrt{(W^2+M^2+Q^2)^2-4W^2 M^2}}\\
&\times\frac{4.5 \times 10^7  (1+\xi)}{Q^6 \xi}.}
The algebraic factors come from the DA and TDA parametrisation (\ced{eq:DACZ}{eq:TDACZ}).
For $\xi=0.8$, we have the $Q^2$ dependence of $\frac{d \sigma}{d\Omega^*_{\pi}}\big|_{\theta^*_\pi=\pi}$ shown in~\cf{fig:dsigma_domega} for $\alpha_s= 0.4$.

\begin{figure}[h]
\includegraphics[height=7cm]{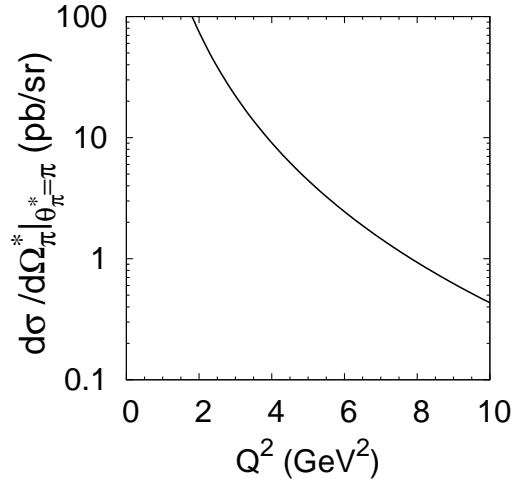}
\caption{$\frac{d \sigma}{d\Omega^*_{\pi}}\big|_{\theta^*_\pi=\pi}$ for $\xi=0.8$ as function of $Q^2$.}
\label{fig:dsigma_domega}
\end{figure}

Our lack of knowledge of the TDAs unfortunately prevents us from comparing our results 
with existing data~\cite{Laveissiere:2003jf}. Indeed these data at $Q^2 = 1 $GeV$^2$ are mostly 
in the resonance region ($W < 1.5$ GeV) except for the large $W$ tail of the distribution, 
which however correspond to small values of the skewedness parameter ($\xi <.3$). We thus 
need a realistic model for the $p\to \pi$ TDAs at smaller values of $\xi$ before discussing 
present data. The issue is more favorable at higher energies, where higher values of $\xi$ 
can be attained above the resonance region, as for instance at HERMES and with CEBAF at 
12 GeV ~\cite{JLAB12}. Our calculation of the cross section, in an admittedly quite narrow 
range of the parameters, can thus serve as a  reasonable input to the feasibility  study of  
backward pion electroproduction at CEBAF at 12 GeV, in the hope to reach the scaling regime, 
in which we are interested.

The corresponding results for the asymptotic choice are three order of magnitude smaller. 
This shows how sensitive  the amplitude is with respect to non-perturbative input of the
DAs. This has to be paralleled with the perfect cancellations in the  proton-form-factor 
calculation in this limit. The breaking of the 
isospin relations for the TDAs prevents some of these cancellations, but the full cross section 
is still shrunk down, whereas the CZ choice gives a much larger contribution as expected.

\section{Conclusion}

Hard-exclusive electroproduction of a meson in the backward region
thus opens a new window in the understanding of hadronic physics in
the  framework of the collinear-factorisation approach of QCD.
Of course, the most important and most difficult problem to solve, 
in order to extract
reliable precise information on the $p \to \pi $ Transition
Distribution Amplitudes from an incomplete set of observables such  
as cross sections and asymmetries,
is to develop a realistic model for the TDAs. This is the subject of 
non-perturbative studies such as, \eg~lattice simulations.
We have derived the limit of these TDAs when the pion momentum 
is small and we have provided a first estimate of the cross section 
in the kinematical regime which should be accessed at JLab at 12 GeV.

This estimate, which is unfortunately reliable only in a restricted kinematical domain,
also shows an interesting sensitivity to the underlying model for the proton DA. 
Beside information about the pion content in protons through the TDAs,
backward pion electroproduction is therefore also likely to bring us information about
the protons DAs themselves. 

Finally, it is worthwhile to note that the analysis presented here could be easily extended
to $ep \to e'n \pi^+$, $ep \to e'\Delta^0 \pi^+$, $ep \to e'\Delta^+ \pi^0$, 
$ep \to e'\Delta^{++} \pi^-$ and similar reactions with a neutron target, 
for which data can also be expected~\cite{private2}.

\acknowledgments

We are thankful to P. Bertin, V. Braun, M. Diehl, M. Gar\c con, F.X. Girod, R. Gothe, M. Guidal, 
C.D. Hyde-Wright, D. Ivanov, 
K. Park, B. Pasquini, A.V. Radyushkin, F. Sabati\'e for useful and stimulating discussions.
This  work  is  supported  by  the
French-Polish scientific agreement Polonium, the Polish grant
1 P03B02828 and the Joint Research
Activity "Generalised Parton Distributions" of the European I3 program
Hadronic Physics, contract RII3-CT-2004-506078. 
L.Sz. is a Visiting Fellow of 
the Fonds National pour la Recherche Scientifique (Belgium). J.P.L is also
a {\it collaborateur scientifique} to the University of Li\`ege and thanks the PTF
group for its hospitality.


\end{document}